\documentclass{Interspeech2024}
\usepackage{amssymb}

\interspeechcameraready

\title{Generalized Source Tracing: Detecting Novel Audio Deepfake Algorithm with Real Emphasis and Fake Dispersion Strategy}

\small \vspace{-10mm}\name[affiliation={1,2}]{Yuankun}{Xie}
\small \name[affiliation={2,*}]{Ruibo}{Fu}
\small \name[affiliation={2}]{Zhengqi}{Wen}
\small \name[affiliation={2,3}]{Zhiyong}{Wang}
\small \name[affiliation={2,3}]{Xiaopeng}{Wang}
\small \name[affiliation={1,*}]{Haonnan}{Cheng}
\small \name[affiliation={1}]{Long}{Ye}
\small \name[affiliation={4}]{Jianhua}{Tao}

\address{
  \small $^1$State Key Laboratory of Media Convergence and Communication, Communication University of China
  \small $^2$Institute of Automation, Chinese Academy of Sciences 
  \small $^3$School of Artificial Intelligence, University of Chinese Academy of Sciences
  \small $^4$Department of Automation and Beijing National Research Center for Information Science and Technology, Tsinghua University}

\email{\{xieyuankun,haonancheng\}@cuc.edu.cn,{ruibo.fu@nlpr.ia.ac.cn.}}

\keywords{audio deepfake algorithm recognition, out-of-distribution detection, audio deepfake detection}

\begin{document}

\maketitle
\renewcommand{\thefootnote}{\fnsymbol{footnote}} 
\footnotetext[1]{Corresponding author.}
\renewcommand{\thefootnote}{\arabic{footnote}}
\begin{abstract}

With the proliferation of deepfake audio, there is an urgent need to investigate their attribution. Current source tracing methods can effectively distinguish in-distribution (ID) categories. However, the rapid evolution of deepfake algorithms poses a critical challenge in the accurate identification of out-of-distribution (OOD) novel deepfake algorithms. In this paper, we propose Real Emphasis and Fake Dispersion (REFD) strategy for audio deepfake algorithm recognition, demonstrating its effectiveness in discriminating ID samples while identifying OOD samples. For effective OOD detection, we first explore current post-hoc OOD methods and propose NSD, a novel OOD approach in identifying novel deepfake algorithms through the similarity consideration of both feature and logits scores. REFD achieves 86.83\% ${F}_{1}$-score as a single system in Audio Deepfake Detection Challenge 2023 Track3, showcasing its state-of-the-art performance. 
\end{abstract}

\section{Introduction}

In recent years, there has been rapid advancement in the field of text-to-speech (TTS) \cite{tan2022naturalspeech,wang2023neural} and voice conversion (VC) \cite{chan2022speechsplit2,tang2022avqvc}, which called deepfake audio. Diverse endeavors and competitions, such as ASVspoof \cite{nautsch2021asvspoof, liu2023asvspoof} and Audio Deepfake Detection challenge \cite{yi2022add,yi2023add}, have been instituted to promote research aimed at developing deepfake countermeasure solutions \cite{xie23c_interspeech}. Current research has demonstrated that in publicly datasets, binary classification tasks of real and fake audio can achieve an Equal Error Rate (EER) around 0.1\% \cite{zhang2023improving}. However, only real/fake classification is not the end. Law enforcement agencies often need to determine the source of deepfake audio for legal rulings. Furthermore, for developers of generative models, it is crucial to trace the source of deepfake audio to protect the intellectual property of their algorithms.  Therefore, it is significant to recognize audio deepfake algorithm.

Recent approaches in the field of Audio Deepfake Algorithm Recognition (ADAR) focus on in-distribution (ID) classification \cite{yan2022initial,zhu2022source,sun2023ai}. However, with the evolution of the deepfake algorithm, distinguishing novel out-of-distribution (OOD) deepfake categories has become increasingly crucial. Recently, a novel challenge in the realm of ADAR, namely Audio Deepfake Detection Challenge 2023 Track3 (ADD2023T3) \cite{yi2023add}, was held to address this issue. This task holds significant importance as it entails not only the detection of diverse fake audio types but also encompasses the presence of unknown generative algorithms during the testing stage. This demands detectors to accurately distinguish categories within the ID while also identifying OOD categories. 

We observed that state-of-the-art approaches \cite{lu2022detecting,qin2022speaker,tian2022deepfake,zeng2022deepfake,wang2022npu} in this track commonly employ multiple classifiers for the multi-classification of real and fake categories using various features and backbones. During testing, OOD methods are utilized to detect unknown classes, and different classifiers scores are fused for a evaluation. We called these one-stage methods.

These one-stage approaches raises some concerns. Firstly, for a classifier, the focus on features differs between distinguishing real and fake and distinguishing among different fake classes. Training on one-stage methods poses a significantly difficulty for the classifier. Secondly, 
in one-stage methods, the genuine class is within ID, which makes it challenging to determine the OOD threshold. In ADD2023T3, OOD samples only generated from the unknown deepfake algorithm. Regarding the ID real class, the OOD data represents a semantic (real or fake) shift. However, for ID fake class, the unknown fake class represents a covariate (fake distribution) shift. Both real and fake class in ID makes it challenging to establish one clear determination threshold for detecting unknown deepfake method.
Lastly, the fusion in the competition, although addressing the different emphases of various features in distinguishing classes, often requires experimenting with different weights on the test set for adaptability, which is inefficient.

\begin{figure*}[!t]
	\centering
	\includegraphics[width= 6in]{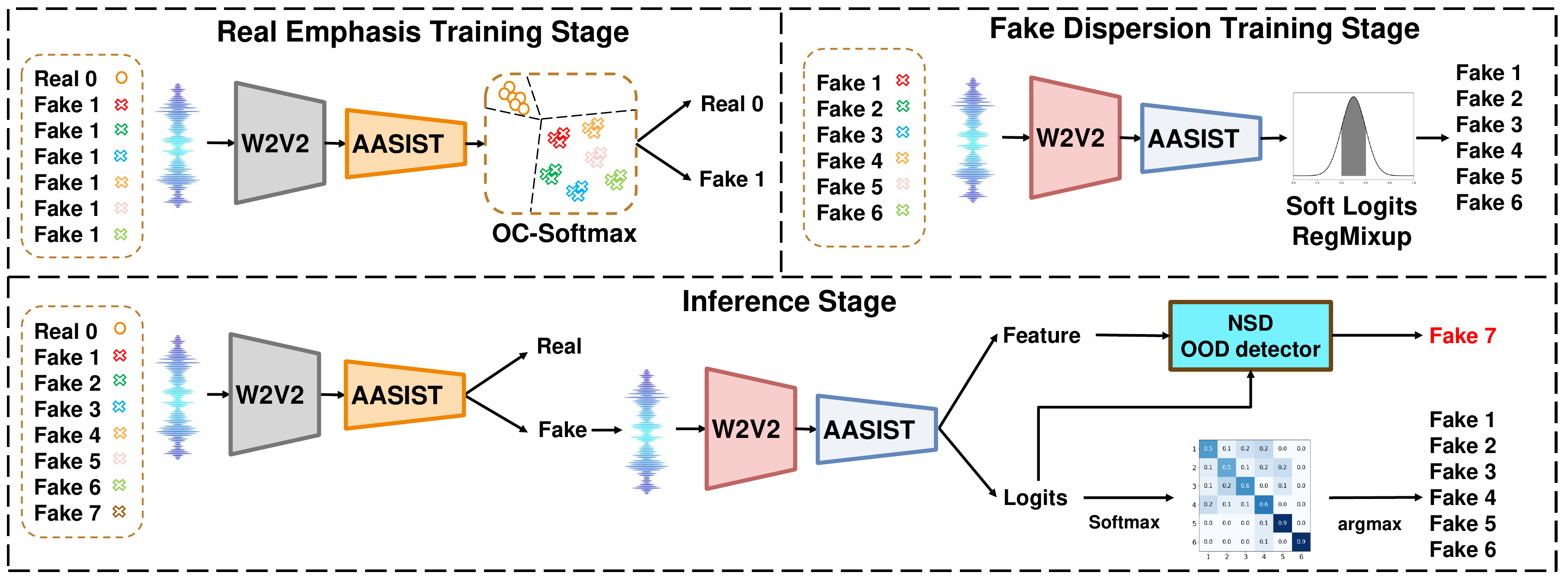}
	\hfil
	\caption{The entire pipeline of our proposed Real Emphasis and Fake Dispersion (REFD) method.}
	\label{fig:pipeline}
\end{figure*}

In this paper, we propose a dual-stage approach for ADAR called Real Emphasis and Fake Dispersion (REFD) strategy. In the Real Emphasis training stage, due to the excellent performance of the current binary classification studies, we continue to adopt a well-established binary classification strategy and incorporate OC-Softmax \cite{zhang2021one} to converge the real decision boundary. At this stage, our primary objective is only to detect the real class, leaving a large isolate feature space for fake classes and unknown fake classes to the second stage. Then, in the Fake Dispersion training stage, our goal is to detect unknown fake classes while classifying ID fake ones. However, the typical cross-entropy with softmax probability often exhibits the issue of overconfidence in classification logits score. This can pose challenges in distinguishing between ID and OOD instances, especially in the application of post-hoc OOD methods, both ID and OOD data may have similarly high output classification logits. To address this, 
we take advantage of Regmixup strategy \cite{pinto2022using}. The Regmixup strategy exhibits strong generalizability inherited from mixup \cite{zhang2017mixup} and maximizes a soft proxy to entropy. This effectively addresses the problem of overconfidence and proves to be effective in post-hoc OOD methods. To detect the unknown audio deepfake algorithm in stage two, we investigated the state-of-the-art post-hoc score based OOD detectors applying in the field of ADAR. Furthermore, we propose a new OOD detection method Novel Similarity Detection (NSD) to effectively detect the OOD data.
NSD determine OOD data by considering both feature similarity and classifier logits, addressing the overconfidence issue associated with a singular viewpoint. This approach helps reduce overconfidence in far-OOD regions while achieving fine-grained detection.

The main contributions of this work are as follows:\begin{itemize}
	\item{We propose Real Emphasis and Fake Dispersion strategy, which can effectively classifying ID samples and detecting OOD samples.}
	\item{We investigated the state-of-the-art post-hoc OOD detectors in the field of ADAR and propose NSD OOD detection method to effective detect the unknown audio deepfake algorithm.}
	\item{Our proposed method was experimentally evaluated on ADD2023T3, demonstrating state-of-the-art performance with an ${F}_{1}$-score of 86.83\% in single system.}
\end{itemize}

\vspace{-3mm}
\section{Method}

The REFD method is shown in Figure \ref{fig:pipeline}. In real emphasis training stage, we concentrate on real class learning and take advantage of OC-Softmax to learn a compact real boundary. In the fake dispersion stage, the RegMixup strategy is employed to learn a soft logits, addressing the overconfidence issue associated with cross-entropy. Lastly, during the inference phase, we employ the real emphasis pre-trained model to detect the genuine class, while the remaining audio is classified using the fake dispersion model. NSD is proposed to to detect the unknown fake class.

\subsection{Backbone}

W2V2-AASIST \cite{tak2022automatic} shows the state-of-the-art performance in the field of audio deepfake detection. However, during training, fine-tuning W2V2 incurs a massive computational burden, especially when applying different losses to the backend features and logits, making training more challenging. Therefore, we freeze the weight of W2V2 and extract the hidden states of W2V2 offline\footnote{https://huggingface.co/facebook/wav2vec2-xls-r-300m} as features and input into the backend AASIST, facilitating training with various losses. 
 
\subsection{Real Emphasis Training Stage}
In real emphasis training stage, all training utterances participate in the training process, with real audio labeled as 0 and fake audio labeled as 1. A two-class prediction logits output (real and fake) is applied in this stage. In this stage, we take advantage of OC-Softmax to learn a one-class compact real decision boundary. Specically, the loss of real emphasis stage (${L}_{RE}$) is defined as follow:
\begin{equation}
	\mathcal{L}_{RE}=\frac{1}{N} \sum_{i=1}^{N} \log \left(1+e^{\alpha\left(m_{y_{i}}-\hat{\boldsymbol{w}}_{0} \hat{\boldsymbol{x}}_{i}\right)(-1)^{y_{i}}}\right),
\end{equation}
where ${x}_{i}$ $\in$ ${R}^{D}$ and $y_{i}$ $\in\{0,1\}$, $w_{0}$ denotes the embedding direction of the target class. The output feature $x$ of AASIST and weight vector $w_{0}$ will be normalized and calculate the cosine similarity of the feature and the target direction. Then, two margins $m_{0}$ $\in\{-1,1\}$, $m_{1}$ $\in\{-1,1\}$ are used to constrain the angle between $x$ and $w_{0}$, denoted as $\theta_{i}$. When $y_{i}=0$, $m_{0}$ is utilized to ensure $\theta_{i}$ is smaller than $\arccos m_{0}$, while for $y_{i}=1$, $m_{1}$ is employed to ensure $\theta_{i}$ is larger than $\arccos m_{1}$. In inference stage, the angle similarity between $x$ and $w_{0}$ is used to determine whether the audio is real or fake.

\subsection{Fake dispersion Training Stage}
The Mixup strategy, built on the fundamentals of Vicinal Risk Minimization (VRM) \cite{chapelle2000vicinal}, synthesizes new samples near the ID distribution. This enrichment of the ID data distribution makes an angular bias between ID and OOD samples. However, employing the Mixup strategy alone often results in limited cross-validation effectiveness and high-entropy behavior \cite{pinto2022using}. Thus, RegMixup \cite{pinto2022using} is proposed which have large cross-validated and maximize a soft proxy to entropy, which is simply combines Empirical Risk Minimization (ERM) \cite{vapnik1991principles} and VRM. The loss of the fake dispersion stage (${L}_{FD}$) with Regmixup strategy is defined as follows:
\begin{equation}
\mathcal{L}_{FD} = \operatorname{CE}\left(p_{\theta}\left(\hat{\mathbf{y}} \mid \mathbf{x}_{i}\right), \mathbf{y}_{i}\right)+\eta \operatorname{CE}\left(p_{\theta}\left(\hat{\mathbf{y}} \mid \overline{\mathbf{x}}_{i}\right), \overline{\mathbf{y}}_{i}\right),
\end{equation}
where for each sample ${x}_{i}$ in a batch, another sample, ${x}_{j}$ is randomly selected from the same batch to obtain the interpolated $\overline{\mathbf{x}}_{i}$ and $\overline{\mathbf{y}}_{i}$, CE denotes the standard cross-entropy loss. 
\begin{table*}[t]
	\caption{${F}_{1}$-score (\%) for different methods in fake dispersion inference stage. }
	\label{tab:fake}
	\centering
	\setlength{\tabcolsep}{1.5mm}{} 
	\renewcommand{\arraystretch}{1.0}
	\begin{tabular}{|c|c|c|c|c|c|c|c|c|c|c|}
		\hline
		Method&OOD &0 &1 &2 &3 &4&5&6&7&AVG\\
		\hline 
		CE (w/o DA)&-&91.73 &52.08 &59.07 &89.90 &96.06 &95.85 &88.16 &0 &71.61 \\ 
		\hline 
		CE &-&91.73 &69.15 &69.98 &96.87 &98.98 &98.40 &93.92 &0 &77.38 \\
		\hline
		CE &NSD&91.73 &67.46 &66.98 &98.11 &98.29 &\bf99.59 &86.86 &43.50 &81.57 \\
		\hline
		CE + Regmixup &-&91.73 &69.08 &79.02 &97.41 &\bf99.33 &96.72 &\bf94.28 &0 &78.45 \\
		\hline
		CE + Regmixup &NSD&91.73 &\bf75.45 &\bf82.32 &\bf98.28 &97.35 &98.42 &92.99 &\bf 58.10 &\bf 86.83 \\
		\hline
	\end{tabular}
\end{table*}
\vspace{-2mm}
\subsection{Inference Stage}
In inference stage, we propose a new OOD detector called Novel Similarity Detection (NSD) to detect novel deepfake algorithm in fake dispersion inference stage. In this stage, lets ${X}_{m}=({x}_{1},...{x}_{m})$ denotes the entire set of $m$ training samples, ${Y}_{n}=({y}_{1},...{y}_{n})$ represents the test samples. We can use the pre-trained fake dispersion model $\phi$ to get the trained feature $\mathbf{Z}_{m}=\phi(||\mathbf{X}_{m}||_{2})$, test feature $\mathbf{T}_{n}=\phi(||\mathbf{Y}_{n}||_{2})$, and their logits ${L}_{m}$ and ${L}_{n}$. NSD treats the entire training domain features as a known class, calculating the cosine similarity between the test domain and training domain features pairwise to obtain the $n \times m$ dimensional similarity matrix ${S}_{NSD}$. However, the normalization used in the calculation of similarity matrix eliminates the dimensional scale of the values, resulting in limited fine-grained detection. Thus, we use the classification logits to scale similarity matrix at the fine-grained value level. The NSD score calculation is as follows:
\begin{equation}
	{S}_{NSD}  =  {T}_{n} * Energy({L}_{n}) \cdot {Z}_{m} * Energy({L}_{m}),
\end{equation}
where Energy \cite{liu2020energy} is the confidence scaling score used to smooth the original logits and enhance generality. We calculate the mean along the $m$ dimensions to obtain the score matrix for $n$ test samples. Samples with scores smaller than the threshold will be identified as novel deepfake algorithms.
\vspace{-2mm}
\section{Experimental Setup}
\subsection{Dataset}
All experiments are conducted on the ADD2023T3 dataset. The ADD2023T3 training set comprises 22,397 audio samples, the development set consists of 8,400 audio samples, both including genuine (class 0) and samples from six different generated methods (class 1-6). The test set contains 79,740 audio samples, including genuine and samples from seven different forgery methods, with one method (class 7) being unknown to the training and development domains.
\begin{table}[h]
	\caption{Number of classes in subsets of ADD2023T3.}
	\label{tab:table1}
	\centering
	\setlength{\tabcolsep}{0.15mm}
	\renewcommand{\arraystretch}{1}
	\begin{tabular}{|c|c|c|c|c|c|c|c|c|c|}
		\hline
		subset&0&1 &2 &3 &4 &5 &6 &7&total\\
		\hline
		train&3200 &3200&3197&3200&3200&3200&3200&0&22397\\
		\hline
		dev&1200 &1200&1200&1200&1200&1200&1200&0&8400\\
		\hline
		eval&9512 &10474&7169&10461&10391&10507&10507&10469&79740\\
		\hline
	\end{tabular}
\end{table}
\vspace{-4mm}
\subsection{Implementation Details}
To alleviate the domain covariate shift between the training and testing domains, we applied the offline data augmentation to the training samples using MUSAN \cite{David2015MUSAN} and RIR \cite{Tom2017A} for five condition. Thus, in the training stage of the REFD method, the training set comprises 111,985 samples in the real emphasis stage and 95,985 in the fake dispersion stage. The development dataset is also augmented to 42,000 samples to simulate a complex scenario. The best-performing model on the development set will be selected for the inference model. The Adam optimizer is adopted with $\beta_{1}= 0.9$, $\beta_{2}= 0.999$, $\varepsilon$ = $10^{-8}$ and weight decay is $10^{-4}$. We train all of the models for 30 epochs. The learning rate is initialized as $10^{-5}$ and halved every 5 epochs. 
\vspace{-3mm}
\section{Results and Analyze}
\subsection{Results in real emphasis inference stage}
In this stage, we focus on the classification accuracy of the real class. We conduct experiments using two backbone, LCNN and AASIST. Experimental results are presented in Table \ref{tab:real}. Regarding backbone, AASIST demonstrated improvements of 3.89\%, 4.21\%, and 5.99\% ${F}_{1}$-score over LCNN in the CE, OC-Softmax, and OC-Softmax-T scenarios, respectively. This led us to adopt AASIST as the backbone network for subsequent experiments. For the loss function, when employing OC-Softmax in inference stage, the threshold will be set to zero. In the case of OC-Softmax-T, we consider only samples with a sufficiently high level of authenticity to be classified as real; the rest are categorized as fake (either existing fake or novel fake). Thus, we set a tight threshold of 0.98 for OC-Softmax-T. Experimental results indicate that ${F}_{1}$-score for W2V2-AASIST with OC-Softmax-T increased by 13.94\% and 6.63\% compared to CE and OC-Softmax, respectively.

\begin{table}[t]
	\caption{${F}_{1}$-score (\%) for the real class in inference phase. CE denotes the original cross-entropy, OC-Softmax-T stands for applying a tight threshold after OC-Softmax.}
	\label{tab:real}
	\centering
	\renewcommand{\arraystretch}{1}
	\setlength{\tabcolsep}{0.15mm}
	\begin{tabular}{|c|c|c|c|}
		\hline
		Model & CE & OC-Softmax & OC-Softmax-T\\
		\hline
		W2V2-LCNN \cite{lavrentyeva2019stc}& 73.90 & 80.89 & 85.74 \\
		\hline
		W2V2-AASIST \cite{tak2022automatic}& 77.79 & 85.10 & \bf 91.73 \\
		\hline
	\end{tabular}
\vspace{-5mm}
\end{table}

\begin{table*}[t]
	\caption{${F}_{1}$-score (\%) for different OOD detector in inference stage. }
	\label{tab:ood}
	\centering
	\setlength{\tabcolsep}{1.5mm}{} 
	\renewcommand{\arraystretch}{0.9}
	\begin{tabular}{|c|c|c|c|c|c|c|c|c|c|}
		\hline
		OOD &0 &1 &2 &3 &4&5&6&7&AVG\\
		\hline
		- &91.73	&69.08	&79.02	&97.41	&99.33	&96.72	&94.28	&0&78.45\\
		\hline
		MSP \cite{hendrycks2016baseline}&91.73 &63.47 &85.19 &98.37 &94.85 &97.96 &80.97 &39.45 &81.50 \\
		\hline
		MaxLogit \cite{hendrycks2019scaling}&91.73 &72.95 &83.13 &98.67 &97.86 &98.45 &89.58 &44.30 &84.58 \\
		\hline
		Energy \cite{liu2020energy}&91.73 &73.49 &81.26 &98.28 &96.15 &98.45 &87.78 &47.56 &84.34 \\
		\hline
		KNN \cite{sun2022out}&91.73 &69.99 &83.26 &98.21 &99.09 &97.44 &91.25 &38.12 &83.64 \\
		\hline 
		Mahalanobis \cite{lee2018simple}&91.73 &64.50 &83.06 &98.31 &98.76 &97.89 &91.03 &45.60 &83.86 \\
		\hline
		NNGuide \cite{park2023nearest}&91.73 &72.44 &82.86 &98.30 &97.49 &98.08 &89.36 &50.05 &85.04 \\
		\hline
		Relation \cite{kim2024neural}&91.73 &70.23 &86.41 &97.94 &96.12 &97.19 &90.38 &51.23 &85.15 \\
		\hline
		NSD &91.73 &75.45 &82.32 &98.28 &97.35 &98.42 &92.99 &58.10 &\bf 86.83 \\
		\hline
	\end{tabular}
\end{table*}
\vspace{-2mm}
\subsection{Results in fake dispersion inference stage}
After inferring real audio in the real emphasis inference stage, the remaining utterances are fed into the pre-trained fake dispersion model to classify into fake categories. For our proposed OOD detector NSD, we determine the threshold for identifying OOD fake classes by setting a threshold on the output logits. The determination of the threshold will be based on achieving the highest average ${F}_{1}$-score. Samples with a score below the threshold will be classified as OOD, while those with a score above the threshold will be classified as ID. For samples classified as ID, the logits will undergo an argmax operation to determine the specific class within the ID category. 

The experimental results are shown in the Table \ref{tab:fake}. In the case of standard CE, NSD method achieves an ${F}_{1}$-score of 43.50\% for class 7, resulting in an overall ${F}_{1}$-score improvement of 4.19\%. Regarding the Regmixup method, although the overall ${F}_{1}$-score increases by only 1.07\% without using NSD, a notable improvement is observed with the application of the NSD method, particularly for OOD class 7, where the score rises to 58.10\%.  This results in an overall ${F}_{1}$-score improvement of 8.38\% compared to non-OOD method, surpassing the 4.19\% improvement achieved with the original CE without the Regmixup strategy. Furthermore, this also implies that the distribution of features and logits obtained through Regmixup strategy are more distinguishable between ID and OOD, rendering it more suitable for the application of post-hoc scored-based OOD methods.
\begin{figure}[!t]
\vspace{-4mm}
	\centering
	\includegraphics[width= 2.0in]{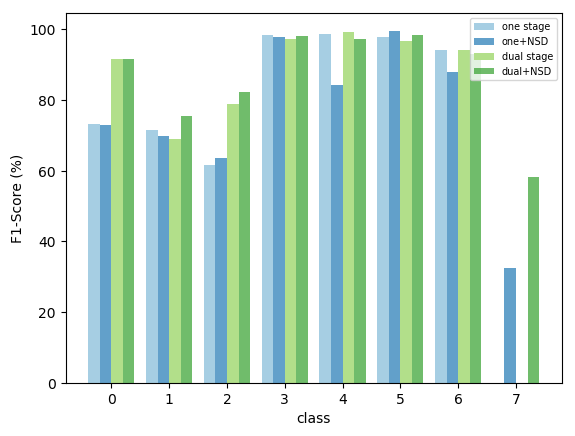}
	\hfil
	\caption{${F}_{1}$-score (\%) comparison between one-stage and dual-stage method.}
	\label{fig:onestage}
\vspace{-6mm}
\end{figure}
\vspace{-2mm}
\subsection{Results for different OOD detectors}
We compared our proposed NSD OOD detector to the state-of-the-art OOD method in Table \ref{tab:ood}. MSP, MaxLogit and Energy are OOD detection methods based on logits scores, while KNN and Mahalanobis are feature distance-based OOD detectors. Relation and NNGuide represent the latest score-based OOD detectors, incorporating both features and logits to jointly establish the OOD threshold. From the results, it can be observed that the final average ${F}_{1}$-score is positively correlated with the scores of OOD classes. Furthermore, relying solely on logits or features to determine the threshold for OOD class does not yield satisfactory results in the field of ADAR. In recent studies, methods like NNGuide and Relation consider both features and logits. NNGuide builds clusters for each ID class based on KNN and determines the threshold by calculating the distance from a test sample to the training cluster. Similarly, Relation calculates the graph relation value for ID classes. Such methods perform well in vision tasks where different ID classes exhibit distinct characteristics. However, in the ADAR domain, variations among different ID classes are often subtle, such as variations in artifact positions. Especially in ADD2023T3, the test set includes noise and other disturbances, making ID differentiation challenging. To address this, we propose the NSD approach, treating all ID categories in the training set as a single category and considering both feature and logits for angle similarity assessment. Experimental results indicate that NSD achieves the highest OOD class ${F}_{1}$-score at 58.10\% and achieve the highest average ${F}_{1}$-score at 86.83\%.
\vspace{-2mm}
\subsection{Compared to one-stage approach}
To validate the effective of the dual-stage approach, we also utilized the CE with Regmixup strategy for overall one-stage training including classes 0-6. During inference, predictions were made directly for classes 0-6, and the NSD method was employed to predict OOD class 7. The experimental results are presented in the Figure \ref{fig:onestage}, highlighting the superior performance of dual-stage method, particularly in the real and OOD classes. In our dual-stage approach REFD, during the real emphasis stage, the classifier concentrate on the differences between real and fake, facilitating the identification of genuine samples. For OOD detection, in the one-stage method, real samples are mixed with ID categories, resulting in a threshold that differs from the fake dispersion stage in the dual-stage approach. This makes it challenging to choose one threshold to distinguish between ID and OOD classes. 
\vspace{-3mm}
\subsection{Compared to state-of-the-art methods}
Table \ref{tab:top5} present the ${F}_{1}$-score of our proposed REFD strategy compared with the top-3 performing methods in ADD2023 Track3: D01 \cite{lu2022detecting}. D02 \cite{qin2022speaker}, D03 \cite{zeng2022deepfake}. As the final scores involve multi-system integration, we also compared the scores of single systems, with individual system results derived from the highest result before score fusion. The results demonstrate that our proposed REFD achieved the highest ${F}_{1}$-score among single systems.

\begin{table}[t]
\vspace{-4mm}
	\caption{${F}_{1}$-score (\%) compared to state-of-the-art methods.}
	\label{tab:top5}
	\centering
	\renewcommand{\arraystretch}{1}
	\setlength{\tabcolsep}{0.3mm}
	\begin{tabular}{|c|c|c|c|c|}
		\hline
		Method & D01 & D02 & D03  & REFD\\
		\hline
		single system & 85.78 & 78.80 & 75.41 &\bf 86.83\\
		\hline
		final result& \bf 89.63 & 83.12& 75.41&86.83\\
		\hline
	\end{tabular}
\vspace{-5mm}
\end{table}
\vspace{-3mm}
\section{Conclusion}
\vspace{-1mm}
This paper propose a dual-stage approach REFD to address the challenge of ADAR. In real emphasis stage, we employ OC-Softmax to identify genuine samples. In fake dispersion stage, we utilize CE with regmixup strategy. This enables us to classify fake samples while generating smooth logits scores, facilitating the application of post-hoc OOD algorithms. Lastly, we investigate state-of-the-art OOD algorithms and propose NSD method, a novel OOD method to detect novel deepfake algorithm. Future work will focus on optimizing the ID feature space by representation learning during the fake dispersion training stage to widen the distribution gap between ID and OOD.

\vspace{-4mm}
\section{Acknowledgements}
This work is supported by the National Natural Science Foundation of China (NSFC) (No.62101553, No.62306316, No.U21B20210, No. 62201571) and in part by the Fundamental Research Funds for the Central Universities under Grant CUC23GZ016.
\vspace{-1mm}

%
%

\bibliographystyle{IEEEtran}
\bibliography{mybib}

\end{document}